\documentclass[conference]{IEEEtran}
\IEEEoverridecommandlockouts
\usepackage{cite}
\usepackage{amsmath,amssymb,amsfonts}
\usepackage{algorithmic}
\usepackage{graphicx}
\usepackage{textcomp}
\usepackage{subcaption}
\usepackage{tabularx}
\usepackage{xcolor}
\usepackage{xspace}
\usepackage[inline]{enumitem} 
\usepackage{booktabs}
\usepackage{multirow}
\usepackage{graphicx} %Loading the package
\usepackage{hyperref}

\usepackage{color, colortbl}
\definecolor{LightCyan}{rgb}{0.88,1,1}

\graphicspath{{img/}}
\begin{document}

\title{Using Structural and Semantic Information to Identify Software Components}
\author{\IEEEauthorblockN{Cezar Sas}
\IEEEauthorblockA{\textit{Bernoulli Institute} \\
\textit{University of Groningen}\\
Groningen, Netherlands \\
c.a.sas@rug.nl}
\and
\IEEEauthorblockN{Andrea Capiluppi}
\IEEEauthorblockA{\textit{Bernoulli Institute}  \\
\textit{University of Groningen} \\
Groningen, Netherlands \\
a.capiluppi@rug.nl}
}

\maketitle

\thispagestyle{plain}
\pagestyle{plain}

\begin{abstract}

Component Based Software Engineering (CBSE) seeks to promote the reuse of software by using existing software modules into the development process. However, the availability of such a reusable component is not immediate and is costly and time consuming. As an alternative, the extraction from pre-existing OO software can be considered.

In this work, we evaluate two community detection algorithms for the task of software components identification.  Considering `components' as `communities', the aim is to evaluate how independent, yet cohesive, the components are when extracted by structurally informed algorithms.

We analyze 412 Java systems and evaluate the cohesion of the extracted communities using four document representation techniques. The evaluation aims to find which algorithm extracts the most semantically cohesive, yet separated communities.

The results show a good performance in both algorithms, however, each has its own strengths. Leiden extracts less cohesive, but better separated, and better clustered components that depend more on similar ones. Infomap, on the other side, creates more cohesive, slightly overlapping clusters that are less likely to depend on other semantically similar components. 

%\paragraph{Conclusion \& Outlook} something like `\textit{the approach is easy to reproduce and can be scaled up to include thousands of systems, and their components}'
%Our approach is easy to reproduce and can be scaled up to many systems for the extraction of components, moreover, it shows the potential of community detection algorithms for the software components identification task

%Large projects are made by many different components, these components TODO.  The dependency between these components means TODO.
%In this work, we focus on how much components that are dependant are semantically similar. 
\end{abstract}


\section{Introduction}

CBSE is an alternative to Object-Oriented (OO) development that aims to make it easier to develop easily reusable and understandable software by using components~\cite{booch1987software}. In order to become useful, CBSE has to produce a repository containing enough components, so that developers do not have to rewrite them from scratch\cite{basha2012CBSE_survey, holmes2008pragmatic}. %, which is an expensive path~\cite{holmes2008pragmatic} 
% An interesting new development for CBSE has been the transition from \textit{monolithic} applications to \textit{microservices}-based architectures~\cite{mazlami2017microservices_graph}.

An alternative to pre-developed and designed components is to extract them from pre-existing software. As an example of this approach, \cite{Chardigny2008ROMANTIC} proposed an architecture recovery approach called ROMANTIC, helped by human experts. Similarly, using particle swarm optimization, \cite{rathee2019mining} proposed a method that uses a meta-heuristic search-based clustering. Frequent usage patterns propelled instead the search heuristics in~\cite{RATHEE2018FUP}.

Other papers, although not necessarily focusing on component reuse, provided algorithms for component extraction: ~\cite{mancordis1998organization, Chiricota2003MQ_clustering} proposed a metric called MQ to measures edge density of graphs, to help in the module selection. The use of graph clustering was often used for the same purpose in later publications~\cite{lee2001coupling, allier2009dynamic}.

The aim of this paper is to evaluate how well-formed the components are when they are automatically extracted by community-detection algorithms~\cite{lancichinetti2009community}. Based on the extracted `communities as modules', we aim to test the semantic cohesiveness of those communities (of features) in the context of the overall software system. As a case study, we use a curated population of 700 Java projects\footnote{\textit{Awesome Java} is a GiHub project that aggregates several hundreds of curated Java projects, available at \url{https://github.com/akullpp/awesome-java}}. After removing empty projects, tutorials, links to websites, or giving issues when analyzing, we obtained a sample of 412 Java projects.

The two objectives of this paper are 
\begin{enumerate*}[label=(\roman*)]
  \item to evaluate the effectiveness of extracting \textit{structure}-informed modules from a software system, and
  \item to detect whether the extracted modules are semantically cohesive.
\end{enumerate*}
In order to assess the first objective, we used two community detection algorithms, \textit{Infomap}~\cite{Rosvall2008map_equation} and \textit{Leiden}~\cite{traag2019leiden}. The second objective is assessed in two ways: firstly, we evaluated whether there is a low cohesion between different modules (e.g., semantic separation); secondly, we checked whether the identified modules are semantically cohesive \textit{per-se}. The source code to replicate our analysis has been made publicly available\footnote{\href{https://github.com/SasCezar/ComponentSemantics/tree/SANER-ERA}{https://github.com/SasCezar/ComponentSemantics/tree/SANER-ERA}}.

% The aim of this paper is to evaluate how well-formed the components are when they are automatically extracted by community-detection algorithms~\cite{lancichinetti2009community}. Based on the extracted `communities as modules', we aim to test the cohesiveness of those communities (of features) in the context of the overall software system. For the stated aim, we analysed a small sample of three OSS Java projects (\textit{antlr4}, \textit{avro}, and \textit{openj9}). Those come from a larger sample\footnote{\textit{Awesome Java} is a GiHub project that aggregates several hundreds of curated Java projects, available at \url{https://github.com/akullpp/awesome-java}} of some 700 Java projects, that will be analyzed as part of the first author's PhD work. 

%\section{Background}
\section{Preliminaries}
\label{sec:_background}
\begin{figure}[htb!]
    \centering
    \includegraphics[width=\columnwidth]{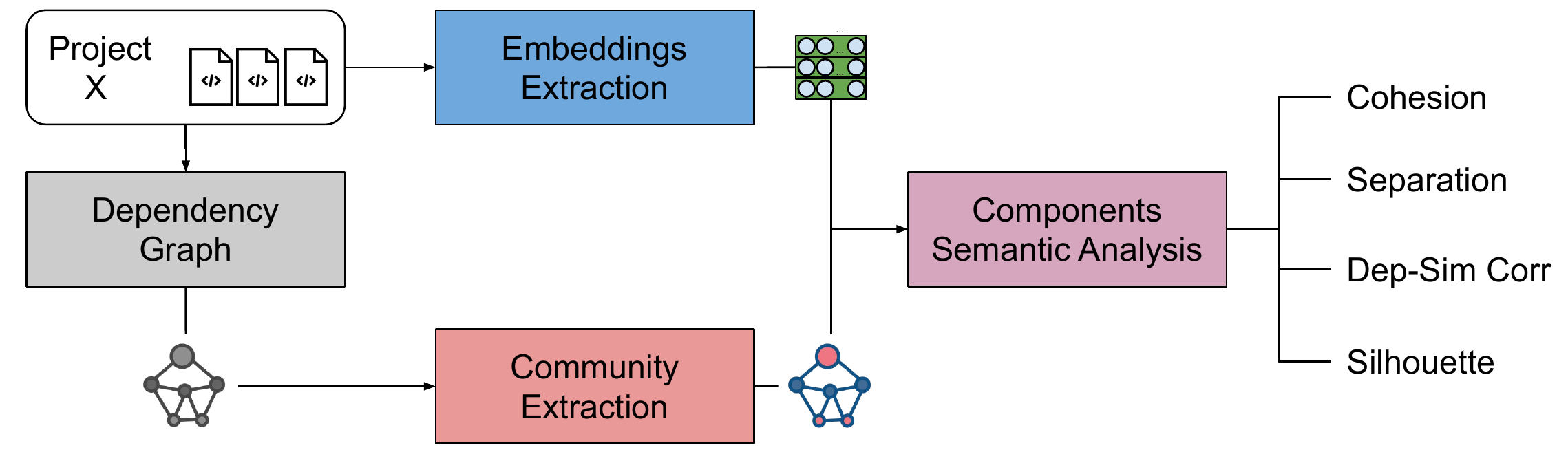}
    \caption{Analysis pipeline. Given a project source code, we extract (a) a vector representation for each file, and (b) the dependency graph. The dependency graph is used as input to the community extraction algorithm. The embeddings and the identified communities are aligned to evaluate the algorithm. }
    \label{fig:pipeline}
\end{figure}

In the following section, we introduce the three steps in this work: how we extracted the structural dependencies (\ref{bkg:_structure}), how we detected the software modules as `communities' (\ref{bkg:_community}) and how we assessed the cohesiveness of the modules (\ref{bkg:_representation}). Figure~\ref{fig:pipeline} presents a visual representation of our pipeline.

\subsection{Structural Dependencies}
\label{bkg:_structure}
The first step of our approach is the extraction of the dependency graph for each project in our sample. Using the Arcan~\cite{fontana2017arcan} tool, we obtained the nodes and the edges describing the dependencies between classes, where the edge weight is the number of uses~\cite{pruijt2017accuracy} of one class in the other. % We obtained the number of vertices and edges described in Table \ref{tab:dataset}.
Arcan parses the abstract syntax tree, extracted with Spoon\cite{pawlak2015spoon}, of the project source files, and it looks for any reference to classes, variables, and methods and saves that information in a graph. 
For example, in the \textit{antlr4} project, we have a dependency between \textit{org.antlr.v4.automata.ParserATNFactory.java} and \textit{org.antlr.v4.tool.LexerGrammar.java} as the former imports the latter. %We also note the weight of such dependency is used by Infomap and Leiden to create the communities.

% \begin{table}[htb!]
% \begin{center}
%     \caption{Size of the analyzed projects in terms of vertices (documents) and edges (dependency between documents).}
%         \begin{tabular}{lccc}
%         \toprule
%             & antlr4    & avro      & openj9    \\ \midrule
% \# Nodes    & 384       & 292       & 910       \\
% \# Edges    & 2,386     & 1,175      & 3,865      \\ \bottomrule
%         \end{tabular}
% \label{tab:dataset}
% \end{center}
% \end{table}

\subsection{Community Detection via Graph Clustering}
\label{bkg:_community}
The second step of our approach uses the relationships identified by the structural dependencies as input for the Community Detection (CD) algorithms. CD (or graph clustering) approaches have been used for various tasks in the software engineering domain besides component identification. Examples include aiding the transition from monolithic applications to microservices architectures~\cite{mazlami2017microservices_graph}, architecture reconstruction~\cite{csora2010software}, and refactoring of software packages~\cite{pan2013refactoring}.

% %We compare two different approaches for community detection (\textbf{how selected?}), the first one is Infomap, and the second is Leiden.
In this step we compare two different types of CD algorithms, Infomap and Leiden: both are popular and effective clustering tools that have been deployed in different domains.

\paragraph{Infomap~\cite{Rosvall2008map_equation}} is an algorithm based on random walks and Huffman coding~\cite{huffman1952coding}. They frame the partitioning of the graph into communities as a code optimization problem. To find an efficient code, Infomap looks for a partition that minimizes the expected description length of a random walk in the graph. In the resulting partition, a community is made of a group of nodes among which information flows quickly and easily.

\paragraph{Leiden~\cite{traag2019leiden}} is an improvement of Louvain~\cite{blondel2008louvain}, with Leiden guaranteeing well connected communities. Both maximise the community modularity by isolating modules with the most dense internal connections (i.e., higher structural cohesion) and the least amount of connections between outside nodes (i.e., lowest structural coupling). Leiden starts by assigning different communities to each vertex. Then, it merges the nodes iteratively, based on the gain in modularity (if there is no gain, the node remains in its current community). The procedure stops when, moving a node into another community, there is no further gain in the modularity.

% Table~\ref{tab:num_components} summarises the number of components (e.g., the communities) as extracted by the Infomap and Leiden algorithms. As visible, Leiden is consistently extracting more (but smaller) components, whereas Infomap can aggregate more classes in the same (larger) component.

% \begin{table}[htb!]
% \begin{center}
%     \caption{Nr of components extracted by Lieden and Infomap}
%         \begin{tabular}{lccc}
%         \toprule
%             & antlr4    & avro  & openj9    \\ \midrule
% Leiden      & 7         & 12    & 26        \\
% Infomap     & 3         & 6     & 16        \\ \bottomrule
%         \end{tabular}
% \label{tab:num_components}
% \end{center}
% \end{table}

\subsection{Semantic Cohesion from Source Code}
\label{bkg:_representation}
The third step of our approach is based on the extraction of representations of project's source code. We used three popular document representation techniques: TF-IDF, BERT, and \textit{fastText}. 

\paragraph{TF-IDF} is a measure that reflects the importance of a word in a document based on a collection of documents. It is the ratio between the number of times the word appears in the document and the amount of information that word provides. %The latter is obtained by dividing the total number of documents by the number of documents containing the term and then taking the logarithm.
TF-IDF does not capture the position in the text, semantics, co-occurrences in different documents, meaning that the created representations are only useful as lexical level features. However, unlike BERT and \textit{fastText}, its vocabulary adapts to the collection. %The final representation of a collection (in our case a software project) is a document-term matrix, where the columns are the terms, and the rows documents.

\paragraph{BERT~\cite{devlin-etal-2019-bert}} is a state-of-the-art neural language model for learning dense vector representation of words (embeddings). It makes use of the Transformer~\cite{vaswani2017transformer}, and learns contextual relations between words in a text. % The vanilla Transformer is based on an encoder-decoder architecture. The encoder reads the input sequence and creates the embedding; the decoder uses these representations to perform the desired task. 
%Compared to other models for word representation, BERT reads the entire input sequence at once, making it a contextual model.
BERT representations are contextual, meaning that word's representation is based on the other words in the sentence (context). %We can obtain an embedding of a sentence or document by taking the vector of the first (special) token. 
In this paper we used the Hugging Face's \textit{bert-base-uncased} pre-trained model~\cite{Wolf2019HuggingFacesTS}. Although this pre-trained model has a vocabulary optimized for natural text, it also adopts a subword segmentation technique called WordPiece~\cite{wu2016wordpiece} (e.g. \textit{playing} becomes \textit{play} $+$ \textit{\#ing}) that aids with out-of-vocabulary words. While this provides additional bits of information, some technical words are nevertheless poorly tokenized (e.g. \textit{servlet} becomes \textit{ser} + \textit{\#v} + \textit{\#let}).

\paragraph{fastText~\cite{bojanowski-etal-2017-enriching}} is a neural embedding model that uses subword information to create word embeddings. \textit{fastText} words are split into \textit{n}-grams, each having its own embedding. The word embedding is an aggregation of the embeddings of all \textit{n}-grams that the word is made of. %It has the advantage that Out-Of-Vocabulary (OOV) words are always represented. 
In contrast to BERT's WordPiece subword segmentation, \textit{fastText}'s \textit{n}-grams have the advantage that technical words are less affected by a segmentation learned from non-technical corpora, due to the smaller size subword information.

%\section{Pre-processing and Document Embeddings}
\section{Dataset}
\label{sec:data}
Our dataset is made by 412 Java projects. Each project has two distinct parts: the structural information from the dependency graph and community detection algorithms, and the semantic information extracted using Natural Language Processing (NLP) techniques.

\subsection{Dependency Graph and Communities}
We extract the dependency graphs using Arcan, and with the previously defined community detection algorithm extract their communities. After extracting the communities, we perform some pre-processing before carrying out our analysis. The structural pre-processing includes the removal of the communities with less than 4 nodes. This choice is driven by the fact that small sized communities skew the result given their high cohesion, usually close to 1, which will advantage algorithms that identify many small modules.

\begin{figure}[htb!]
    \centering
    \includegraphics[width=0.9\columnwidth]{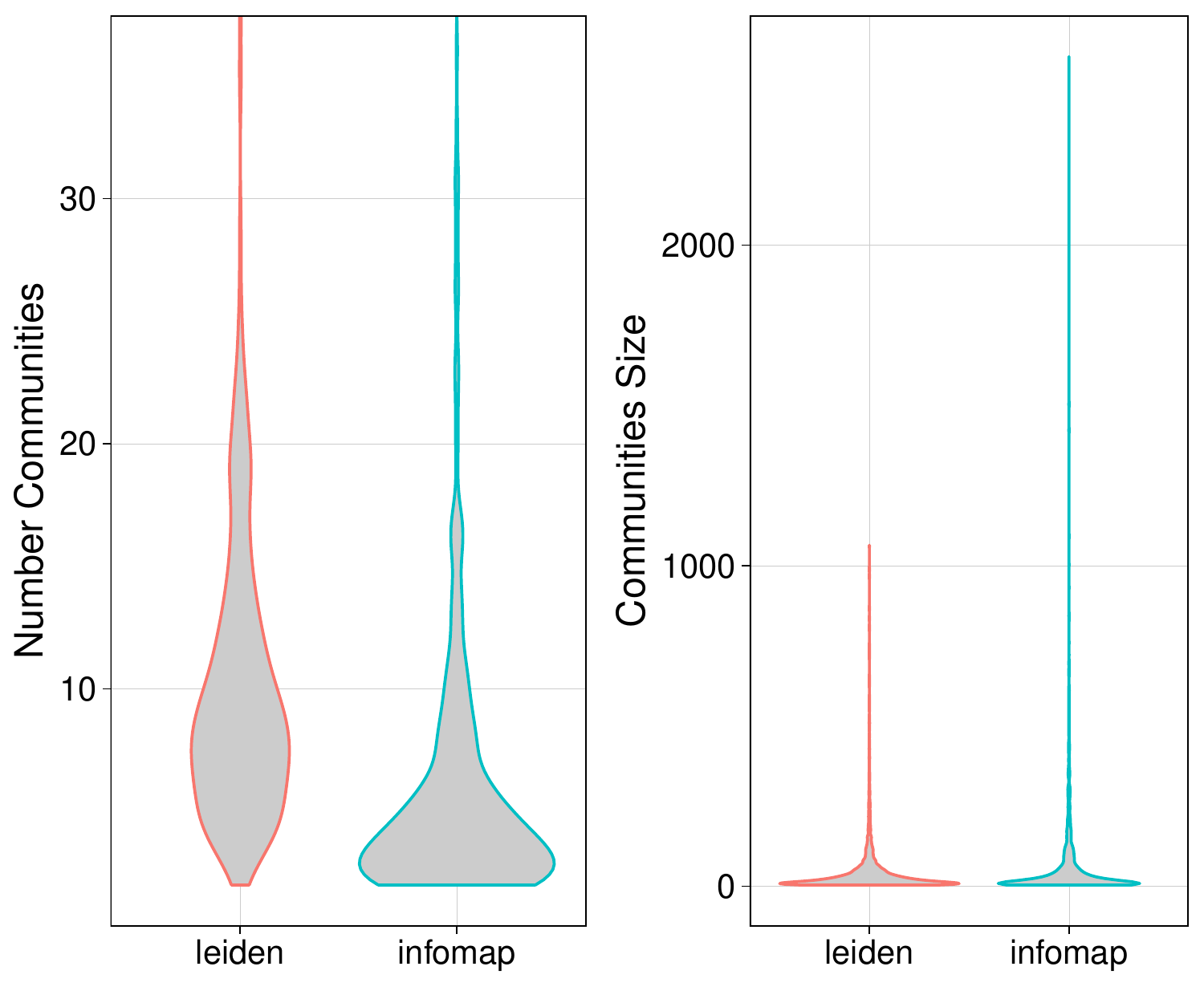}
    \caption{Distribution of the number (left) and size (right) of the identified communities.}
    \label{fig:comm_distribution}
\end{figure}

After the removal of the smaller communities we obtain the results in Figure~\ref{fig:comm_distribution} which summarises the distribution of the number and size of identified components. Infomap has fewer communities, while their size is larger with more outliers. Leiden extracts a larger number of smaller communities, and it has fewer outliers. 

\subsection{Document Embeddings}
The extraction of features representing the documents is performed on two conceptual levels, package- and class-names, and source code. We create four different representation:

\paragraph{Embeddings from package- and class-names} The first feature extraction uses BERT to perform an embedding of the cleaned text of the \textit{package} and \textit{class name}. The cleaning consists of splitting the package name by dot; next, we remove the first two elements of the list as they represent the organization that developed the software. The final step is to split the camel case strings into words and remove any Java keyword. For example \textit{org.antlr.v4.tool.LexerGrammar} becomes \textit{``v4 tool lexer grammar"}.  %which is used as input to BERT to create the embedding.%. The obtained sequence is used as input sentence to BERT which creates its embedding.

\paragraph{BERT embeddings from source code} The second feature extraction method also uses BERT; in this case, we perform the embedding on the set of \textit{identifiers} extracted from the source code file. For the extraction of the identifiers contained in the source code, we used the \textit{tree-sitter} parser generator tool\footnote{\href{https://github.com/tree-sitter/tree-sitter}{https://github.com/tree-sitter/tree-sitter}}. It makes easy to get the identifiers, without keywords, from the annotated concrete syntax tree created using a grammar for Java code. We clean the identifiers as before, and remove common Java terms that do not add much semantically (e.g., `\textit{main}', `\textit{println}', etc). %Figure \ref{fig:doc2_sample} shows an example of the preprocessing.

\paragraph{fastText embeddings from source code} The third document representation solution focuses on the identifiers again, but it uses \textit{fastText} as the embedding model. For our experiments we use the \textit{wiki.en} %\footnote{\url{https://dl.fbaipublicfiles.com/fasttext/vectors-wiki/wiki.en.vec}}
pre-trained model.

\paragraph{TF-IDF bag of words from source code} The last feature extraction method uses TF-IDF for the creation of the document embedding. To reduce the vocabulary, and normalize words, we use the lemma of the identifiers as our tokens. Furthermore, for each of the analysed systems, we limit the vocabulary size to the top $1,000$ terms (the mean vocabulary size in our dataset is around $1,200$ tokens). %We consider as the document collection for the creation of the document-term matrix the source code files of the project. We limit the vocabulary size to the top $1000$ terms.

\section{Experiments and Metrics}
\label{sec:exp_res}
%We performed experiments to %STATE WHAT WE ARE LOOKING FOR.
%evaluate the quality of the extracted components.
Below we describe the metrics that were used to evaluate Infomap and Leiden in terms of semantic cohesion of components and separation between different components. 

\subsection{Semantic Cohesion}
\label{sec:_exp_semcohe}
We first evaluate the internal semantic cohesion of the components using the cosine similarity. This should indicate how cohesive are the components found by the CD algorithm. %Given two vectors, the cosine similarity ranges between $+1$, and $-1$. With $+1$ meaning that the two vectors have the same orientation, $-1$ that their meaning is opposite, and $0$ that they are orthogonal.

The pairwise similarity is evaluated between all the vectors describing the nodes in the community and then averaged. Then, for each project, we average the internal similarities of all communities. Formally, we compute it as follows:

\begin{equation*}
coh = \frac{1}{N}\sum_{k}^{N}\left(\frac{2}{|C_k|\cdot(|C_k|-1)}\sum_{i,j \land i < j}^{|C_k|}sim(\textbf{x}_{k,i}, \textbf{x}_{k,j})\right)
    \label{eq:cohesion}
\end{equation*}

where $N$ is the number of communities, $|C_k|$ is the cardinality of community $k$ (with at least 4 nodes), $sim$ is the cosine similarity measure, and  $\textbf{x}_{k,i}$ is the $i$-th node of community $k$.

% The average of the components mean similarity for each project is presented in Table \ref{tab:cohesion}. We can notice that there is a large difference in the similarity scores between BERT and TF-IDF. This is also present in the other analysis. This is due to the different type of representation obtained by the two methods, TF-IDF is sparse and only syntactic, making it more difficult for the latter to obtain high similarity scores for documents that contain different words with similar meaning. 

%The analysis shows that the components extracted from both Leiden and Infomap have a high cohesion with each of the three features. While both have a high cohesion, Infomap gives more cohesive components 5 out of 9 times. 
% These results alone do not give the full picture of how well the algorithms performed, we also need to check the semantic separation between components.

%also, C3 is a metric for cohesion: https://www.ijsr.net/get_abstract.php?paper_id=20131221
% which other ones could we use?

\subsection{Semantic Separation}
\label{sec:_semasepa}
The extracted components should have low semantic similarity between each other. This is measured by the cosine similarity between the representing vectors of each community. Since each component is composed of many nodes, each with its own embedding, the community vector is obtained aggregating in a single vector all the information using the mean function. More formally:

\begin{equation*}
sep = \frac{2}{N\cdot(N-1)}\sum_{i,j \land i < j}^{N}sim(\textbf{c}_i, \textbf{c}_j)%;  \textbf{c}_i = \frac{1}{|C_i|}\sum_{j}^{|C_i|}\textbf{x}_{i,j}
\end{equation*}

where $\textbf{c}_i$ is the representation of the $i$-th community:

\begin{equation*}
    \textbf{c}_i = \frac{1}{|C_i|}\sum_{j}^{|C_i|}\textbf{x}_{i,j}
\end{equation*}

\subsection{Silhouette}
\label{sec:silhouette}
The average similarity scores above do not give a global idea of how well the nodes are grouped together semantically.  We can get a combined measure of both the internal cohesion and external separation by using the Silhouette coefficient\cite{ROUSSEEUW198753}. This is a measure of how similar an object is to its own cluster (cohesion), as compared to other clusters (separation). It uses the information from the nodes to get an average quality score. %The values range from $+1$ (i.e., perfect cluster assignment) to $-1$ (i.e., each element is assigned to the wrong cluster). A value of $0$ indicates overlapping clusters. 
For our experiments we use the cosine distance. %, defined as $1 - cos\_similarity$.

% The silhouette results are presented in Table \ref{tab:silhouette}. The results give a better overview of the previous analysis. We can see that Leiden gets consistently better results than Infomap (8 times). This is caused by more semantically sparse components as it can also be noticed in Figure~\ref{subfig:TSNE_infomap}.

\subsection{Correlation Between Dependencies and Similarity}
\label{sec:corr_sim_dep}
The final analysis measures the correlation between the number of dependencies connecting the components and their semantic similarity. 

We would like this score to be slightly negative, since having similar components with a high dependency between each other might indicate \begin{enumerate*}[label=(\roman*)]\item that the algorithm splits a component into two, or \item that there are different components that perform different parts of the same tasks.\end{enumerate*}

For this experiment, we measured the number of edges between each community in a pairwise fashion. We consider the edges as undirected and use the total amount of dependencies between communities. %Moreover, we measured the semantic similarity between components, and, as for the semantic separation, we used the mean of the embeddings of each vector as the component representation. 
Using the previously defined features, we measured the semantic similarity between components. We used the mean of the embeddings of each vector as the component representation. The resulting representations are used to compute the pairwise similarities. % We also compared the two embedding levels: package, and document.
We compute the correlation of the two variables using Pearson's $r$.

% The results for each project are shown in Table \ref{tab:pearson}. For both Leiden and Infomap there is no significant correlation between the semantic similarity of components and their level of dependencies. However, Infomap has two negative values, even if overall it is lower only 4 times, and only on the smaller projects. This is caused by it having a large component with many nodes that span across a wide semantic space.

\section{Results and Discussion}
The results of our analysis are presented in Table~\ref{tab:results}. They are aggregated, showing how many times either Infomap or Leiden is better than the other on the specific metric. % We also evaluated the \textit{`Cohesion $>$ Separation (\%)'} scenario to measure the number times the cohesion is higher than the separation in a project. % Instead, Table~\ref{tab:agreement} we can see how.

\begin{figure}[htb!]
    \centering
    \includegraphics[width=0.9\columnwidth]{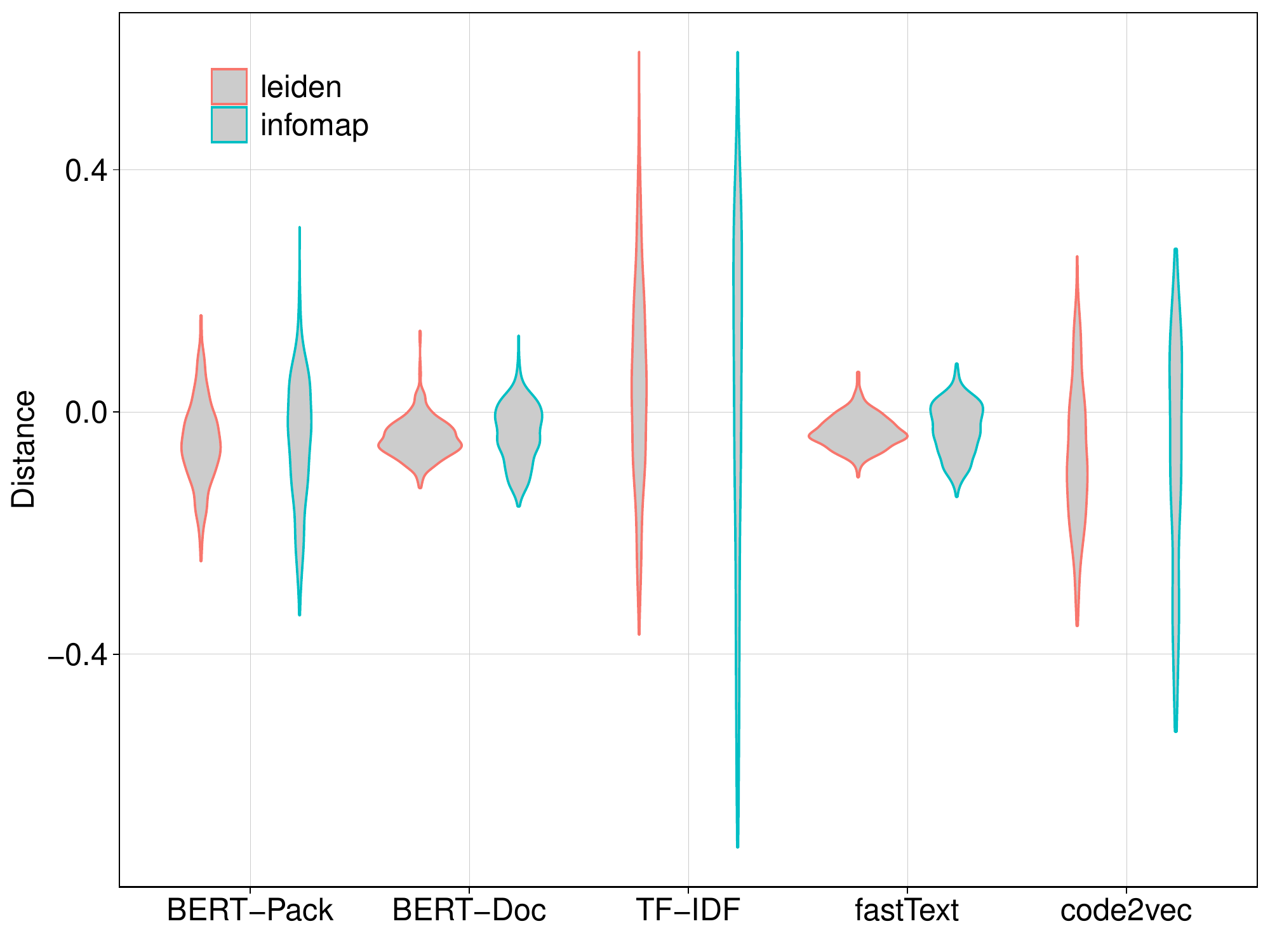}
    \caption{Distribution of the difference between cohesion and separation scores. Leiden is less scattered, however, it also has more negative scores.}
    \label{fig:distances}
\end{figure}

We can notice that Infomap has in general a higher cohesion across all features, while Leiden has a better separation between components. However, the \textit{`Cohesion $>$ Separation (\%)'}, which measures the number of times (and the percentage) the cohesion is higher than the separation in a project, shows that although Leiden has better results when using TF-IDF, Infomap performs better on the other three methods. Figure~\ref{fig:distances} shows the distribution of the differences.

\begin{table*}[htb!]
\begin{center}
    \caption{Aggregated results: numbers represent the \# of projects in the sample. \textbf{Bold} highlights the better algorithm.}
        \begin{tabularx}{2\columnwidth}{lXXXXXXXX}
        \toprule
        & \multicolumn{4}{c}{BERT} & \multicolumn{2}{c}{\multirow{3}{*}{TF-IDF}}& \multicolumn{2}{c}{\multirow{3}{*}{fastText}} \\
        \cmidrule(lr){2-5}
        \multirow{1}{*}{Metric} & \multicolumn{2}{c}{Package} & \multicolumn{2}{c}{Document} &  \\
        \cmidrule(lr){2-3}\cmidrule(lr){4-5} \cmidrule(lr){6-7}\cmidrule(lr){8-9}
                  &  Leiden     & Infomap     & Leiden    & Infomap   & Leiden       & Infomap   & Leiden       & Infomap  \\ \midrule
        {Cohesion}  & $183$     & $\textbf{229}$   & $172$  & $\textbf{240}$  & $170$     & $\textbf{240}$ & $177$ & $\textbf{235}$   \\
        {Separation}    & $\textbf{223}$     & $189$   & $\textbf{208}$  & $204$ & $\textbf{292}$ & $190$  & $\textbf{217}$     & $195$    \\

        {Silhouette}  & $\textbf{256}$     & $156$   & $\textbf{251}$  & $161$  & $\textbf{323}$     & $89$  & $\textbf{234}$ & $178$  \\
        {Dep-Sim Corr}  & $86$     & $\textbf{326}$   & $98$  & $\textbf{314}$  & $90$     & $\textbf{322}$  & $104$ & $\textbf{308}$ \\ 
        
        \rowcolor{LightCyan}
        {Cohesion $>$ Separation (\%)}  & $85 \;(20\%)$     & $\textbf{140 \;(33\%)}$   & $35 \;(8\%)$  & $\textbf{105 \;(25\%)}$  & $\textbf{247 \;(60\%)}$     & $222 \;(53\%)$  & $48 \;(11\%)$ & $\textbf{120 \;(29\%)}$ \\\bottomrule
        \end{tabularx}
\label{tab:results}
\end{center}
\end{table*}

\begin{figure*}[htb!]
\subfloat[Leiden]{%
  \includegraphics[clip,width=\columnwidth]{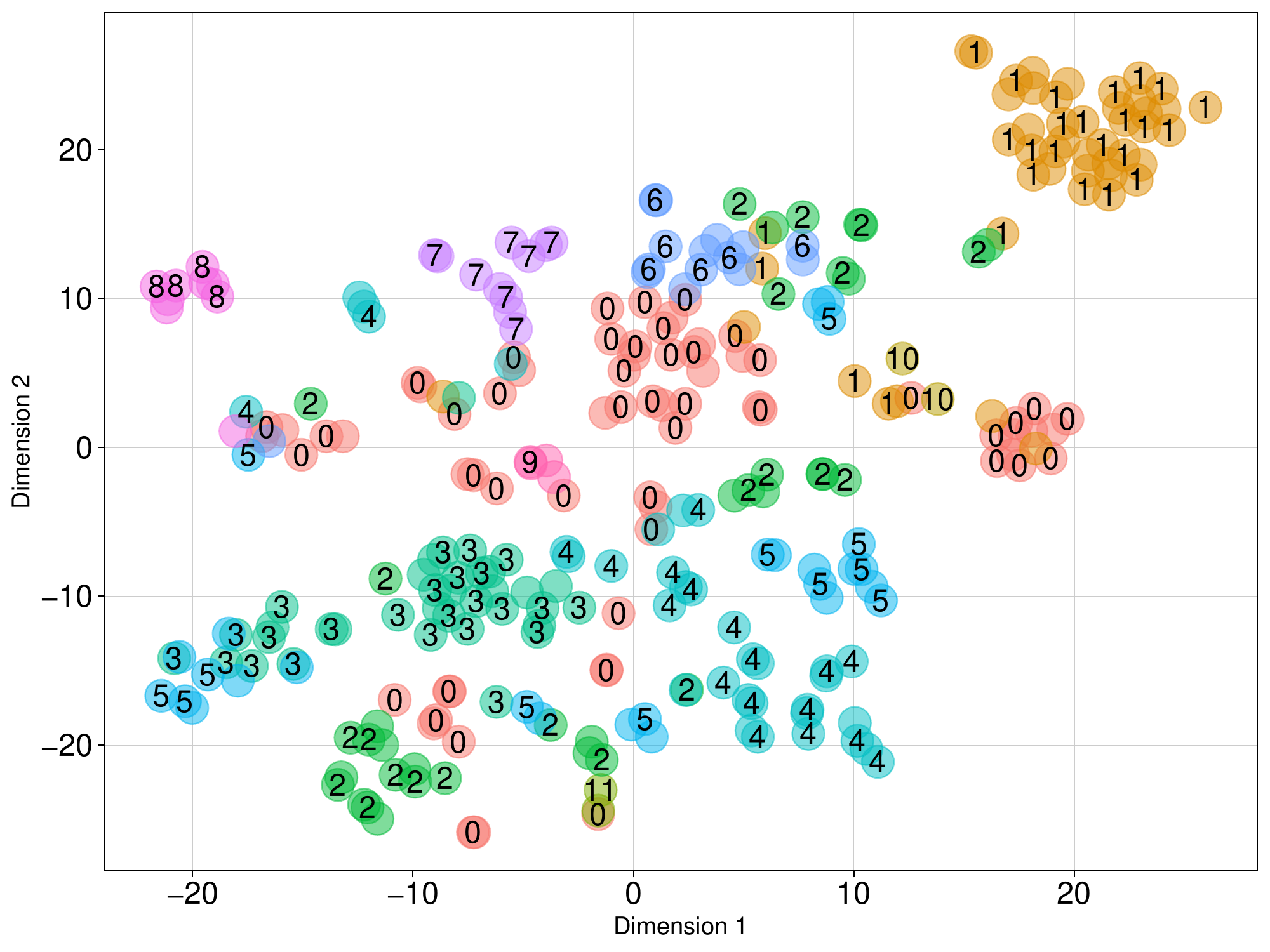}
\label{subfig:TSNE_leiden}
}
\hspace{0.2em}
\subfloat[Infomap]{%
  \includegraphics[clip,width=\columnwidth]{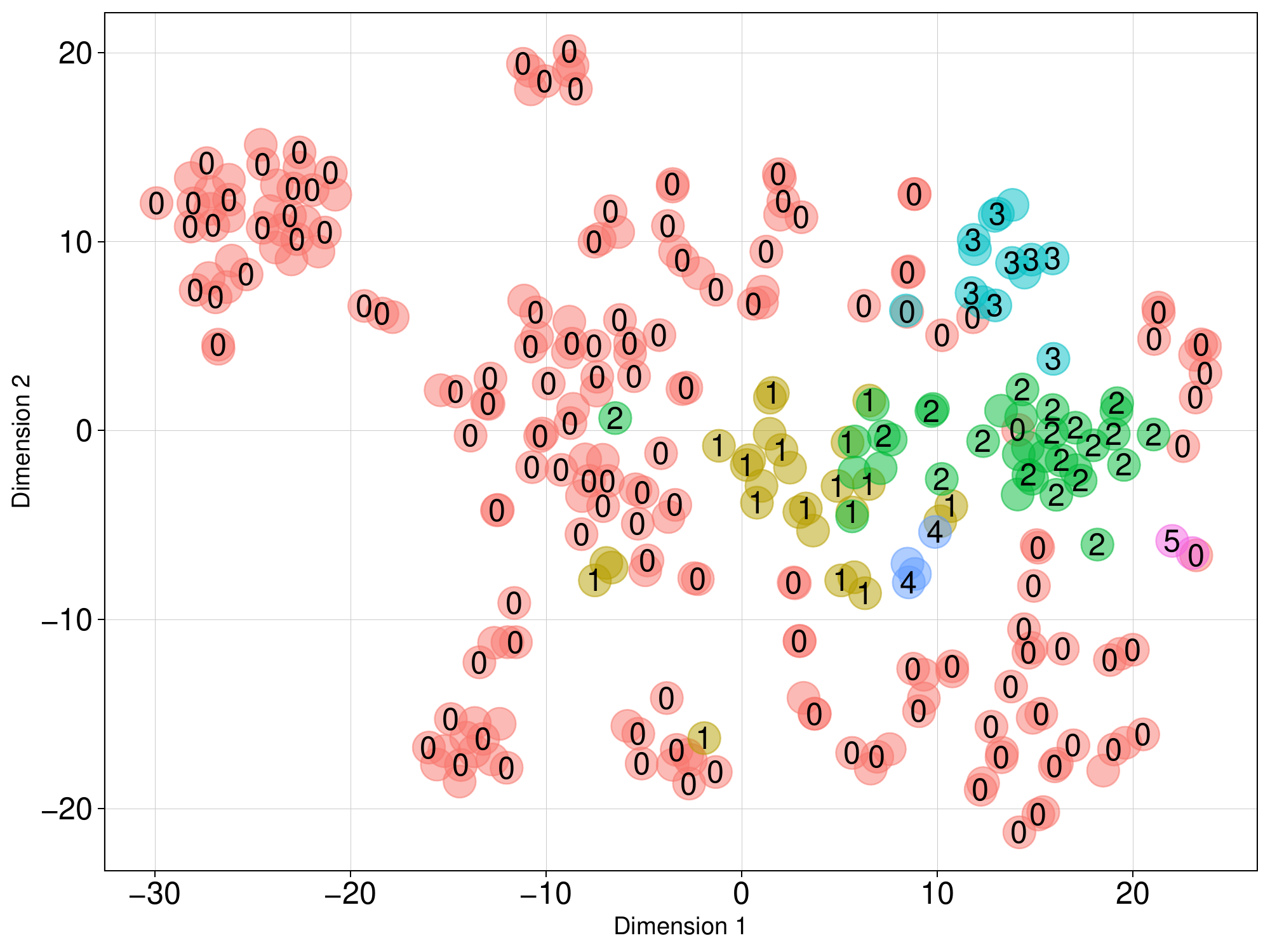}
  \label{subfig:TSNE_infomap}
}
\caption{\textit{avro}'s TF-IDF features reduced to two dimensions. The spatial position of the node represents its semantic, the color (and label) the component it belongs to. Leiden's communities 2, 5, and 0 are scattered, while the others are more cohesive. In Infomap community 0 spans the entire semantic space, while communities 3, 1, and 2 are cohesive.}% The communities with same number are not the same between the two algorithm.}
\label{fig:TSNE}
\end{figure*}

Leiden shows better scores using the Silhouette metric: this result is actually due to the fact that Infomap communities are larger in size, but sparser in terms of semantic. This is reflected by nodes clustered in one large component, but that are semantically closer to other components (see Figure~\ref{fig:TSNE}). 

Lastly, we have the results of the correlation between the similarity of the identified modules, and their dependency. In this experiment, Leiden is the better algorithm in around $75\%$ of the cases. However, this score can be influenced by having few very large communities and many smaller ones. 

While the communities extracted by Infomap can become too large, this could be fixed by running a \textit{multi-level} extraction: the algorithm could be semi-automatically forced to optimize its first partitioning by running over the individual communities again. However, this will cause the opposite issue on smaller graphs: this has to be further investigated in order to find the best policy given the project size. %identifying communities.

A threat to our methodology is the removal of the small communities. While the choice might seem arbitrary, it also has to be considered that, beside skewing the results, components with less than 4 nodes are unlikely to be of much interest in large projects. 
%%%%%%  IS the following part clear?:  %%%%%%%
Furthermore, the averaged representations and similarities, and the small differences between the internal and external similarity scores, give only a preliminary idea of what the algorithms really identify. Lastly, the analyses are dependant on the chosen representation technique: however, this threat is reduced by having four different representation.

\section{Conclusion and Future Work}
\label{sec:conclusion}
In this work we compared two different community detection algorithms (Infomap and Leiden) for identifying software components. We evaluated the algorithms by using four different representation methods. The results show that Leiden extracts more components with fewer nodes. Its components are less cohesive, yet better separated, and their nodes are better clustered, but the components depend more on similar ones. Infomap, on the other hand, creates fewer but bigger components that are more cohesive, with slightly overlapping clusters and are less likely to depend more on semantically similar components. 

As future work, we plan to qualitatively evaluate each of the extracted components individually, and to assess possible automatic refinements to the content of the components based on their semantics. Furthermore, components should be assigned a topic to make them easily understandable and accessible.
Lastly, we plan to evaluate other algorithms that use both the structural and semantic information to assess if they achieve better results and consistency across the different metrics.

\bibliography{references}

\end{document}